# Engineering Educators' Perspectives on the Impact of Generative AI in Higher Education

Umama Dewan
*Department of Computer Science George Mason University* Fairfax, USA
0009-0000-5192-261X

Ashish Hingle
*Department of Information Sciences and Technology*
*George Mason University*
Fairfax, USA
0000-0002-6178-1256

Nora McDonald
*Department of Information Sciences and Technology*
*George Mason University*
Fairfax, USA
0000-0003-0216-5573

Aditya Johri
*Department of Information Sciences and Technology*
*George Mason University*
Fairfax, USA
johri@gmu.edu

*Abstract*—The introduction of generative artificial intelligence (GenAI) has been met with a mix of reactions by higher education institutions, ranging from consternation and resistance to wholehearted acceptance. Previous work has looked at the discourse and policies adopted by universities across the U.S. as well as educators, along with the inclusion of GenAI-related content and topics in higher education. Building on previous research, this study reports findings from a survey of engineering educators on their use of and perspectives toward generative AI. Specifically, we surveyed 98 educators from engineering, computer science, and education who participated in a workshop on GenAI in Engineering Education to learn about their perspectives on using these tools for teaching and research. We asked them about their use of and comfort with GenAI, their overall perspectives on GenAI, the challenges and potential harms of using it for teaching, learning, and research, and examined whether their approach to using and integrating GenAI in their classroom influenced their experiences with GenAI and perceptions of it. Consistent with other research in GenAI education, we found that while the majority of participants were somewhat familiar with GenAI, reported use varied considerably. We found that educators harbored mostly hopeful and positive views about the potential of GenAI. We also found that those who engaged more with their students on the topic of GenAI, both as *communicators* (those who spoke directly with their students) and as *incorporators* (those who included it in their syllabus), tend to be more positive about its contribution to learning, while also being more attuned to its potential abuses. These findings suggest that integrating and engaging with generative AI is essential to foster productive interactions between instructors and students around this technology. Our work ultimately contributes to the evolving discourse on GenAI use, integration, and avoidance within educational settings. Through exploratory quantitative research, we have identified specific areas for further investigation.

## I. INTRODUCTION

The introduction of GenAI tools, especially the release of ChatGPT, has led to increasing concerns about their use, especially within higher education. Generative AI technologies can produce responses and code so similar to human outputs that they may easily be mistaken for them, raising concerns about their implications for academic integrity, the originality of student work, and the evolving role of educators in guiding learning [1]–[3]. Many educators and academic institutions are currently facing challenges in determining how best to incorporate these tools into their teaching methods without compromising educational values [4]. Key issues include the potential for plagiarism, questions about the trustworthiness of AI-generated content, and concerns that reliance on these tools might weaken students' critical thinking and problem-solving abilities [5], [6]. As GenAI use continues to grow in educational settings, it is essential to develop clear guidelines and policies to safeguard the standards and integrity of higher education [7]. This research aims to understand educators' views on the advantages and potential pitfalls of using GenAI in teaching, learning, and research.

To better understand university-level educators' perspectives on the usage and impact of GenAI, as well as how their actions regarding GenAI usage influence their perspectives and practices, we conducted an empirical study surveying practitioners, including faculty, post-docs, and graduate students, in higher education institutions, with a focus on those from engineering and computing backgrounds. By capturing insights from educators and those closely involved in designing and implementing coursework, activities, and tools, we aim to explore how these technologies are currently being integrated into course curricula and research, the benefits and challenges associated with their use, and the broader implications of GenAI for teaching, learning, and research. Our study aims to answer the following research questions:

- **RQ1:** What are educators' perspectives on the use of GenAI in higher education for teaching and research?
- **RQ2.** What challenges and potential harms do they perceive with the use of GenAI in teaching and research?
- **RQ3:** How does their approach towards using GenAI influence their perceptions and experiences with GenAI?

## II. RELEVANT WORK

### A. The Impact of AI in Education

The role of AI in education has been a key topic of discussion among researchers, practitioners, and administrators, and its importance has grown steadily over the past decade [8], [9].

One driving factor is the increasing integration of innovative digital technologies into various educational elements, such as personalized assistance through intelligent tutoring systems [10], conducting assessment [11], and AI-enabled learning management systems (LMS) [12]. Furthermore, the ongoing adaptation of tools and forms of pedagogy has made AI more accessible in different age groups [8]. The impact on education has been extensive and will only continue as the tools improve, and new technology is employed to solve problems [13], [14]. One key inflection point marking a substantial increase in AI use and interest in education was the release of *OpenAI's ChatGPT* in late 2022. ChatGPT, a GenAI tool, became popular because it combined a conversational agent interface with a powerful large language model (LLM), making it easy for the general public to start using it. While the tool initially provided sometimes incorrect or superficial responses, it encouraged experimentation across various domains and sparked other companies to strategize their own application of GenAI [15]. Students quickly adopted GenAI, prompting some institutions to ban its use in classrooms [16], [17], while others appear willing to embrace the technology by providing licenses for students and faculty [18].

As research on GenAI continues to expand, its significant impact on different elements of the educational ecosystem is becoming more apparent. Educators may leverage GenAI to support their teaching by identifying patterns across students' backgrounds, preparedness, and motivation levels. The technology could serve as an "early warning system" for identifying students who may be struggling [13]. Additionally, innovative applications of GenAI are emerging to support adaptive pedagogy. For example, Abolnejadian et al. [19] developed a custom learning platform with GenAI that offers personalized educational materials tailored to student's backgrounds. Educators also use GenAI tools to build solutions, modify content and teaching processes [20], and address student needs efficiently and directly [19].

From a student perspective, GenAI can provide personalized and interactive instruction [19], [21] and adaptive learning environments and experiences without an instructor having to curate every turn [22], [23]. But reliance on GenAI can also alter the help-seeking behaviors of students, and the quality of their experiences [24]. There is an increasing body of research looking at student trust of GenAI [25] as well as the impact on teacher-student relationships. For instance, instructors' use of GenAI can sometimes present challenges for transparency and exacerbate power imbalances, further undermining trust [26], [27].

### B. Educator's Perspectives of GenAI

Several surveys have reported on educators' perspectives of GenAI across fields and tasks. These surveys often frame their inquiries through contrasting viewpoints, portraying the technology as either full of potential or fraught with challenges [13], [14], [28], or categorizing educators' opinions as seeing GenAI as either a helpful tool or a potential threat to education [29]. It is also evident that educators struggle with the unrealized potential of GenAI, often reporting that they use it primarily for superficial tasks, while lamenting the lack of institutional support for its effective and ethical use [14].

In their survey of teachers who had used GenAI at least once, Kaplan-Rakowski et al. [28] found generally positive experiences with GenAI tools and noted more frequent use corresponded with increasingly positive perspectives.

An important theme across these studies is the experimental nature of educator's engagements with GenAI, frequently testing its capabilities by observing how it handles their assignments and assessments [3], [4], [14], [28].

Prior work has shown that, as with any technology, the adoption within higher education is a non-linear process, with marked differences in use by early adopters, likely adopters, and non-adopters [30]. For institutions planning to accelerate the process of adoption, a better understanding of factors influencing these differences is essential. In our study, we bring this differentiated understanding by further dividing our sample across dimensions of adoption: educators who communicate to the class about GenAI and those who do not, and educators who incorporate GenAI use in their course syllabus and those who do not. So far as we know, no research has looked at the relationship between instructors' experiences through the use of GenAI, their communications with students about GenAI, and their perspectives on its influence on education.

## III. METHODOLOGY

### A. Survey Design

The survey aimed to gather broader insights on how GenAI tools are perceived and utilized within academic settings. The survey was hosted on the *Qualtrics* platform and comprised 13 questions intended to capture both quantitative and qualitative data. The questions can broadly be categorized into three types: demographic information, perspectives on and use of GenAI in both teaching and research, and the potential benefits and harms perceived by educators regarding the use of GenAI. Our survey included multiple choice and Likert scale questions to measure attitudes and perceptions regarding GenAI, as well as open-ended questions that encouraged participants to elaborate on their views.

### B. Participant Demographics

We disseminated this survey to 160 educators who joined a workshop on GenAI in engineering education, yielding a total of 98 responses. Participants in the study came from an academic background, with the sample comprising 26% teaching faculty, 24% tenured professors, 18% tenure-track professors at the assistant and associate levels, 13% graduate students, and 4% post-doctoral candidates.

| Higher Education Position | Percentage |
|---|---|
| Tenured professor | 24.5% |
| Associate tenure-track professor | 9.2% |
| Assistant tenure-track professor | 9.2% |
| Teaching faculty | 25.5% |
| Post-doctoral candidate | 4.1% |
| Graduate student | 13.3% |
| Other | 14.3% |

TABLE I
PARTICIPANTS REPORTED ACADEMIC POSITION

Over two-thirds (70%) of our participants came from the engineering field, with an additional 7% from computer science and information technology. The remaining participants represented fields such as education, library and information science, geosciences education, and management (Table II). Years of professional experience are well-represented across different levels, ranging from 1–5 years to over 20 years (Table III).

| Professional Field | Percentage |
|---|---|
| Engineering (any discipline) | 70.4% |
| Computer Science (including IT) | 7.1% |
| Education | 12.2% |
| Other | 10.2% |

TABLE II
PARTICIPANTS REPORTED PROFESSIONAL FIELD

| Years of Professional Experience | Percentage |
|---|---|
| 1-5 years | 28% |
| 5-10 years | 18% |
| 10-20 years | 24% |
| 20+ years | 30% |

TABLE III
PARTICIPANTS REPORTED TIME IN THE FIELD

### C. Data Analysis

Our analysis followed a structured and systematic approach to uncover both quantitative trends and qualitative insights regarding the use and impact of GenAI and potential risks associated with its use in higher education.

We organized the quantitative data from the multiple-choice and Likert scale questions, which provided numerical insights into the participants' perceptions of GenAI's role in education. Responses were exported from the Qualtrics platform into Excel spreadsheets, where descriptive statistics such as frequency distributions and percentages were calculated. These statistics helped to identify overarching patterns, such as the percentage of participants who viewed GenAI as having a positive, neutral, or negative impact on various aspects of higher education, including course conduction, student engagement, and academic integrity. The analysis of this quantitative data provided a broad overview of trends in participants' responses.

We manually open-coded our free-response survey questions [31]. One researcher reviewed the responses and developed a codebook to categorize the emerging themes. Some of the codes related to potential risks and harms were informed by existing literature [32], ensuring that our analysis built on previously established research while allowing for the development of new themes.

### IV. FINDINGS

#### A. Educators' Perspectives on GenAI

*1) Familiarity with GenAI:* Of our participants, 78% reported being somewhat or very familiar with GenAI. In contrast, 17% indicated that they were somewhat unfamiliar, and 4% stated that they were not at all familiar with GenAI (Table IV).

| Personal Experience with GenAI | Percentage |
|---|---|
| Very familiar | 18.3% |
| Somewhat familiar | 60.2% |
| Somewhat unfamiliar | 17.2% |
| Not at all familiar | 4.3% |

TABLE IV
PARTICIPANTS' REPORTED FAMILIARITY WITH GENAI

*2) Perception about GenAI:* Most participants held a positive view of GenAI's impact on higher education and their professional practices. For instance, 77% of respondents agreed, either strongly or somewhat, that GenAI will transform higher education classrooms for good, with a similar percentage agreeing that it will influence how they design their curriculum. This suggests a broad consensus among educators about the potential for GenAI to reshape teaching methodologies and course structure.

In the context of the professional engineering workplace, an even larger portion of participants (83%) agreed that GenAI will have a transformative impact on the professional engineering workplace for good. Furthermore, 80% respondents acknowledged that GenAI will transform their course preparation and grading processes, while 70% respondents believed it will significantly impact their research. These figures highlight a widespread recognition of GenAI's potential across both academic and professional settings, with a particular emphasis on its role in transforming educational and research workflows (Figure 1).

*3) Impact of GenAI in different course activities:* GenAI has the most pronounced positive impact in the areas of code proofreading, code writing, and the generation of new code. Specifically, 47% of respondents noted a positive influence of GenAI on these tasks, highlighting its utility in enhancing the accuracy, efficiency, and quality of coding processes.

Moreover, 40% of participants reported a positive impact of GenAI in learning course concepts. Another domain identified as positively impacted by GenAI is writing, with 36% of respondents indicating a positive influence (Figure 2).

*4) Use of GenAI in courses:* About half of the participants (54%) mentioned that they include policies about GenAI in

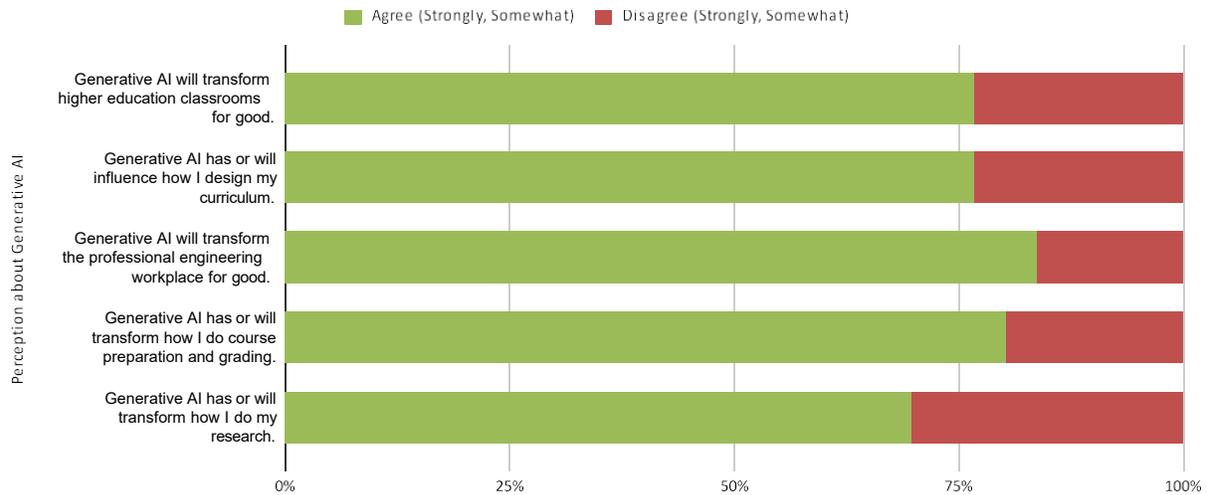

Fig. 1. Overall Perception about GenAI

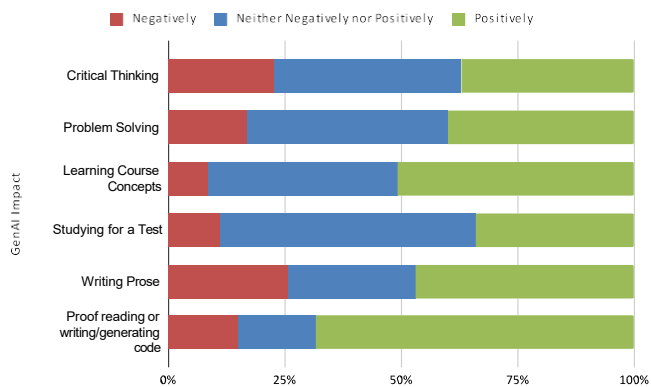

Fig. 2. Impact of GenAI on different course activities

their syllabus. Fewer, a little over a third (40%) have actually incorporated GenAI into their course assignments.

We asked educators to indicate their agreement with statements addressing their use of GenAI, beliefs about student policy violations involving GenAI, and the availability of resources to address such violations (see Figure 3). While these statements do not fully capture the nuances of their views, they are grounded in literature on uses and concerns.

The large majority of educators (73%) specified that they do not have experience with or believe their students are violating their policies about the use of GenAI in their course. At the same time, the majority of people do not feel they have the tools they need to deal with student use of GenAI against their stated syllabus policy (60%) but regularly discuss the ethics of using it with their students (66%) (Figure 3).

*5) Ways GenAI is assisting teaching:* GenAI is being utilized in various aspects of course preparation, with assignments and assessments being the most prominent areas. 65% of the participants reported using GenAI to prepare assignments, while 55% of the participants used it for preparing assessments. This suggests that educators are finding GenAI useful in creating and designing tasks that test students' knowledge and understanding.

| Ways GenAI is assisting teaching | Percentage |
|---|---|
| Prepare syllabus | 33.7% |
| Prepare course content | 53.0% |
| Prepare assessments | 55.4% |
| Prepare assignments | 65.1% |

TABLE V
APPLICATIONS OF GENAI IN ASSISTING TEACHING

Moreover, half of the participants (53%) are using AI to prepare course content, indicating its role in shaping the overall structure and material of the courses. However, fewer participants (34%) are using it for syllabus preparation.

*6) Ways GenAI is assisting research:* GenAI is being used in various aspects of research, with writing being the most prominent area (74%). Half of the participants report using GenAI for data analysis and 40% are using it for research design. Fewer participants (17%) reported using AI for data generation and other unspecified research tasks (20%).

| Ways GenAI is assisting research | Percentage |
|---|---|
| Data generation | 16.8% |
| Data analysis | 54.5% |
| Research design | 40.3% |
| Writing | 74.0% |
| Other | 19.5% |

TABLE VI
APPLICATIONS OF GENAI IN ASSISTING RESEARCH

*B. Potential challenges and harms associated with the use of GenAI*

To examine the potential risks and harms of GenAI as perceived by educators, we incorporated open-ended questions

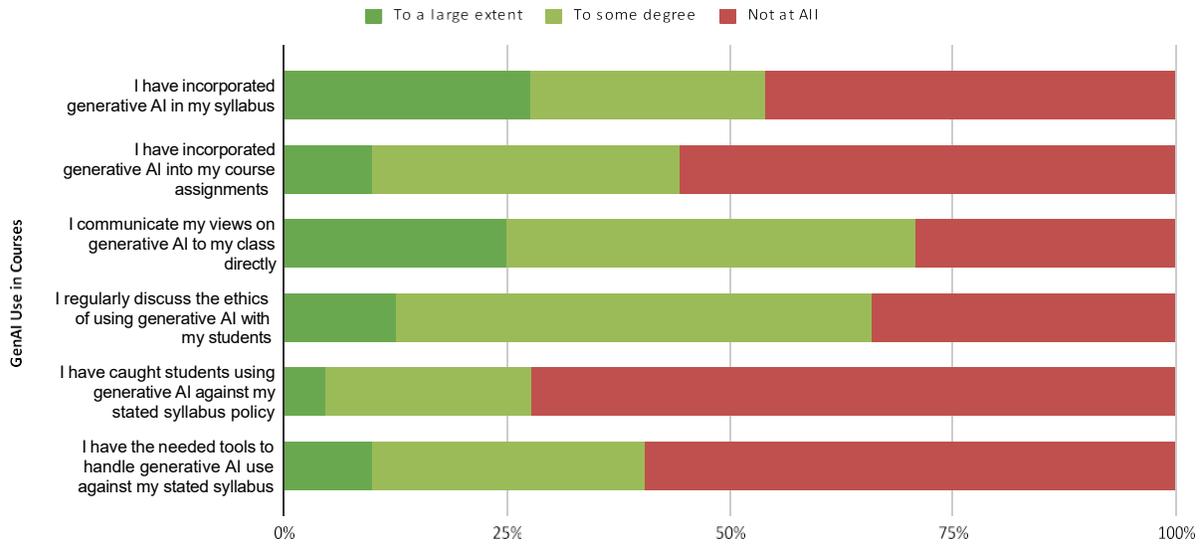

Fig. 3. Use of GenAI in different course activities

into our survey. Table VII shows the codes and their definitions.

*1) Challenges foreseen with GenAI in teaching:* Table VIII highlights several challenges that educators anticipate regarding the use of GenAI in teaching, as indicated in their open-ended responses. A total of 68 participants complete an open-ended response about teaching. Privacy and awareness emerge as the most significant concern, with almost a third (31%) highlighting this issue, reflecting a widespread apprehension about the security and ethical implications of AI tools in education. Hindering learning and difficulty with fair assessments are equally notable challenges, each mentioned by 22% of participants providing responses to this question. Meanwhile, the veracity of GenAI output (12%) and lack of understanding of AI (8%) are concerns among a few participants.

*2) Challenges foreseen with GenAI in research:* Table IX highlights key challenges that educators foresee with the use of GenAI in research, as indicated in their open-ended responses. A total of 64 participants gave open-ends about challenges of using GenAI in research. The veracity of GenAI output stands out as the most significant concern, with 23% participants expressing doubts about the accuracy and reliability of AI-generated content in research contexts. This is closely followed by plagiarism, cited by 19% participants, which reflects concerns about the potential misuse of AI tools to generate content without proper attribution or originality. Privacy and awareness are notably minor concerns, mentioned by only 3% participants. This suggests that research priorities are primarily focused on ensuring the integrity and quality of AI-generated outputs, with comparatively less emphasis on broader impacts, such as learning outcomes and content depth.

*3) Specific potential risks or harms related to GenAI:* Table X presents the risks that educators spontaneously report regarding the use of GenAI in education. A total of 63 participants provided open-ends about the potential risks and harms foreseen with the use of GenAI in education. The most frequently cited concern is the potential of GenAI to hinder learning (40%). Similarly, privacy and awareness issues were highlighted by 37% of participants, while 32% expressed doubts about the accuracy of GenAI outputs. Surprisingly, only 11% mentioned plagiarism. Lack of understanding was cited by 8% of participants.

*C. The impact of educators' approaches on their perceptions and experiences with GenAI*

Ali et al. [32] identified exemplary practices among R1 Engineering and Computing educators, including presenting authentic experiences and taking a transparent and thorough approach to articulating GenAI policies. As part of our analysis, we identified educators who were taking similar approaches to understand how that related to their perceptions and experiences with GenAI. These included:

- **GenAI Communicators:** Educators who communicate their views on GenAI directly to their class. This group includes those who answered "to some degree" or "to a large extent" to "I communicate my views on GenAI to my class directly." (N=50, 67%) as opposed to those who answered "not at all" to the same statement (N=25, 33%).
- **GenAI Incorporators in Syllabus:** Educators who incorporate GenAI in their syllabus. This group includes those who answered "to some degree" or "to a large extent" to "I have incorporated GenAI in my syllabus." (N=48, 54%) as opposed to those who answered "not at all" to the same statement (N=41, 46%).

*1) GenAI Communicators:* Those who directly communicate their views about GenAI (we refer to as *GenAI Communicators*) feel more comfortable with the tools available (50% of communicators vs 15% of non-communicators), and are more likely to incorporate GenAI into their syllabus (72% of communicators vs 19% of non-communicators). These instructors also regularly discuss GenAI ethics with students (70% of communicators vs 16% of non-communicators). Interestingly, these educators are also more likely to have caught their students using GenAI in ways that are against their stated syllabus (34% of communicators vs 4% of non-communicators). Table XI summarizes these findings.

| Codes | Definitions |
|---|---|
| Hinders Learning | Use of GenAI tools may be detrimental to learning experience, building of foundational skills, possibly lead to difficulties in future careers, limiting critical thinking etc. |
| Privacy and Awareness | Legal, privacy, security, and ethical implications of using GenAI |
| Veracity of GenAI Output | Whether GenAI output is accurate, biased, or misleading |
| Difficulty with fair assessments | Not being able to provide fair assessments to student works due to the use of GenAI |
| Lack of understanding | Not being able to understand the usage criteria of GenAI |
| Plagiarism | No acknowledgment, annotations, informal citation, or formal citation (i.e., APA, MLA, Chicago, etc.) of the content generated by GenAI |
| Surface-level writing | Writing done by GenAI lacks depth |

TABLE VII
CODES AND DEFINITIONS

| Challenges foreseen with GenAI in Teaching | Count |
|---|---|
| Hinders learning | 22.1% |
| Privacy and awareness | 30.9% |
| Veracity of GenAI output | 11.7% |
| Difficulty with fair assessment | 22.1% |
| Lack of understanding | 8.8% |

TABLE VIII
CHALLENGES FORESEEN WITH GENAI IN TEACHING

| Challenges foreseen with GenAI in Research | Count |
|---|---|
| Privacy and awareness | 3% |
| Veracity of GenAI output | 23.4% |
| Plagiarism | 18.8% |

TABLE IX
CHALLENGES FORESEEN WITH GENAI IN RESEARCH

While communicators and non-communicators generally agree on GenAI's potential to transform classrooms and its role in higher education, there seems to be a relationship between discussing GenAI and holding more positive perceptions of its impact. For example, communicators consistently show higher agreement rates across all categories of perception regarding the transformative impact of GenAI. This may suggest that educators who are more open to discussing and learning about GenAI are more optimistic about its benefits. Higher disagreement levels in the non-communicators group, particularly regarding curriculum design and research impact, could indicate uncertainty or lack of exposure to GenAI's potential. It might also suggest that non-communicators have more reservations or lack confidence in their understanding or application of GenAI. Table XII illustrates these data.

| Potential Risks or Harms | Count |
|---|---|
| Hinders learning | 39.7% |
| Privacy and awareness | 36.5% |
| Veracity of GenAI output | 31.7% |
| Plagiarism | 11.1% |
| Lack of understanding | 7.9% |

TABLE X
POTENTIAL RISKS OR HARMS FORESEEN WITH GENAI

When it comes to specific activities, communicators are more positive about the role of GenAI with respect to critical thinking, learning concepts, and writing. For example, in "Learning Course Concepts," 50% of communicators see a positive impact, while only 28% of non-communicators share this view. Similarly, in "Proofreading or Writing/Generating Code," 56% of communicators perceive a positive impact, compared to 28% of the other group. It can be seen from Table XIII that, across most educational activities, a higher percentage of communicator participants perceive a positive impact of GenAI compared to non-communicator participants.

*2) GenAI Incorporators in Syllabus:* Educators who incorporate GenAI into their syllabus (Incorporators) demonstrate higher engagement and preparedness in dealing with GenAI-related issues compared to those who do not (Non-Incorporators). These educators are more likely to communicate their views on GenAI directly to their students (90% of Incorporators vs. 49% of Non-Incorporators) and feel more confident in having the necessary tools to handle GenAI use against their syllabus policies (50% of Incorporators vs. 29% of Non-Incorporators). Moreover, Incorporators educators are more proactive in discussing the ethics of using GenAI, with 92% engaging in these discussions compared to only 35% of Non-Incorporators educators. In addition, while 54% of Incorporators educators have integrated GenAI into their course assignments to some extent, only 24% of Non-Incorporators educators have done so. This indicates a greater willingness among Incorporators educators to explore and incorporate new technologies in their teaching practices, potentially enriching the learning experience. Table XIV summarizes these findings.

Although both Incorporators and Non-Incorporators generally agree on the transformative potential of GenAI in higher education, Non-Incorporators surprisingly display greater confidence in its positive impact. This trend is evident in their perceptions across various aspects of GenAI integration. For example, while 55% of Incorporators somewhat agree that GenAI will transform higher education classrooms for good, a greater percentage of Non-Incorporators (64%) somewhat agree with this statement. Similarly, Incorporators show a nuanced perspective when it comes to curriculum design, with 70% expressing agreement (40% somewhat agree, 30% strongly agree) about GenAI's influence on shaping their courses. In contrast, 85% of Non-Incorporators agree that GenAI will impact curriculum design, with a higher proportion (54%) strongly agreeing. This indicates that Non-Incorporators, who may be less directly engaged with GenAI, are still optimistic about its role in future educational frameworks. Their stronger agreement might reflect a theoretical appreciation of GenAI's potential, without the practical challenges that Incorporators might experience in integrating these technologies into their courses. Table XV illustrates these data.

| GenAI Use | Group | Response | | |
|---|---|---|---|---|
| | | Not at all | To some degree | To a large extent |
| I have incorporated generative AI in my syllabus. | Communicators | 28% | 46% | 26% |
| | Non-Communicators | 81% | 4% | 15% |
| I feel that I have the tools I need to deal with use of generative AI against my stated syllabus policy. | Communicators | 50% | 14% | 36% |
| | Non-Communicators | 85% | 0% | 15% |
| I regularly discuss the ethics of using generative AI with my students | Communicators | 10% | 20% | 70% |
| | Non-Communicators | 84% | 0% | 16% |
| I have caught students using generative AI against my stated syllabus policy | Communicators | 58% | 8% | 34% |
| | Non-Communicators | 96% | 0% | 4% |
| I have incorporated generative AI into my course assignments. | Communicators | 46% | 6% | 48% |
| | Non-Communicators | 81% | 0% | 19% |

TABLE XI
GENAI USE: COMPARISON BETWEEN COMMUNICATORS AND NON-COMMUNICATORS

| Perception Statement | Group | Response | | | |
|---|---|---|---|---|---|
| | | Strongly Disagree | Somewhat Disagree | Somewhat Agree | Strongly Agree |
| Generative AI will transform higher education classrooms for good. | Communicators | 6% | 8% | 59% | 27% |
| | Non-Communicators | 4% | 32% | 56% | 8% |
| Generative AI has or will influence how I design my curriculum. | Communicators | 0% | 16% | 43% | 41% |
| | Non-Communicators | 12% | 16% | 52% | 20% |
| Generative AI will transform the professional engineering workplace for good. | Communicators | 4% | 4% | 59% | 33% |
| | Non-Communicators | 0% | 24% | 56% | 20% |
| Generative AI has or will transform how I do course preparation and grading. | Communicators | 0% | 10% | 51% | 39% |
| | Non-Communicators | 4% | 28% | 44% | 24% |
| Generative AI has or will transform how I do my research. | Communicators | 4% | 16% | 39% | 41% |
| | Non-Communicators | 0% | 48% | 28% | 24% |

TABLE XII
PERCEPTIONS ABOUT GENAI: COMPARISON BETWEEN COMMUNICATORS AND NON-COMMUNICATORS

Interestingly, while acknowledging some benefits of GenAI in specific areas like problem-solving and learning concepts, Incorporators are more cautious about its broader educational impact. Non-Incorporators, on the other hand, exhibit more variability in their responses, with stronger positive perceptions in areas like critical thinking and code-related tasks but also a higher tendency to see GenAI as irrelevant to their teaching practices. This indicates differing levels of familiarity and acceptance of GenAI, highlighting the need for further exploration and support to address these variations in perception and usage. Table XVI summarizes these findings.

While *GenAI Incorporators* and *GenAI Communicators* represent two distinct but often overlapping groups, each highlights different dimensions of educators' engagement with GenAI. The overlap between these groups illustrates that many educators who incorporate GenAI into their courses also tend to communicate their perspectives openly, but there are important nuances. For example, while communicators may excel at fostering dialogue and shaping student attitudes, incorporators are more focused on the structural integration of GenAI within curricula. This distinction is critical as it reveals different approaches to leveraging GenAI: one oriented around fostering understanding and ethical considerations, and the other emphasizing direct pedagogical application.

Examining both groups allows us to better understand the multifaceted ways educators engage with GenAI and the implications for classroom dynamics, student engagement, and policy-making. This dual focus underscores the importance of strategies that combine communication, transparency, and practical implementation to maximize GenAI's potential while addressing its challenges in educational settings.

V. DISCUSSION

Our work contributes to the ongoing narrative about GenAI use, integration, and avoidance in educational settings. Through exploratory quantitative research, we have identified segments for future testing, which we elaborate on below.

*A. GenAI Integration*

We found that instructors who engage directly with students about GenAI and incorporate it into their syllabus tend to feel more comfortable with the technology, particularly with the tools available (table XI and XIV). These instructors also have a slightly more positive outlook on GenAI across various academic activities, even if all instructors seem generally positive about the technology (table XIII and XVI). We also found that by integrating GenAI into their curriculum, they may be more proactive in discussing ethical considerations, fostering a responsible approach to the technology. Moreover, those who readily communicate and integrate GenAI into their syllabus seem more realistic about potential misuse and (we speculate) may even take steps to modify their syllabus to address this risk (table XII and XV).

| GenAI Impact | Group | Response | | | |
|---|---|---|---|---|---|
| | | Negatively (Somewhat, Very) | Neither Negatively nor Positively | Positively (Somewhat, Very) | Does not apply |
| Critical thinking | Communicators | 24% | 34% | 30% | 12% |
| | Non-Communicators | 13% | 26% | 26% | 35% |
| Problem solving | Communicators | 18% | 36% | 40% | 6% |
| | Non-Communicators | 8% | 36% | 12% | 44% |
| Learning course concepts | Communicators | 10% | 34% | 50% | 6% |
| | Non-Communicators | 4% | 24% | 28% | 44% |
| Studying for a test | Communicators | 8% | 38% | 22% | 32% |
| | Non-Communicators | 8% | 24% | 12% | 56% |
| Writing prose | Communicators | 22% | 22% | 42% | 14% |
| | Non-Communicators | 12% | 16% | 28% | 44% |
| Proof Reading or Writing/generating Code | Communicators | 12% | 12% | 56% | 20% |
| | Non-Communicators | 8% | 12% | 28% | 52% |

TABLE XIII
GENAI IMPACT ON COURSE ACTIVITIES: COMPARISON BETWEEN COMMUNICATORS AND NON-COMMUNICATORS

| GenAI Use | Group | Response | | |
|---|---|---|---|---|
| | | Not at all | To some degree | To a large extent |
| I communicate my views on generative AI to my class directly. | Incorporators | 10% | 48% | 42% |
| | Non-Incorporators | 51% | 44% | 5% |
| I feel that I have the tools I need to deal with use of generative AI against my stated syllabus policy. | Incorporators | 50% | 38% | 12% |
| | Non-Incorporators | 71% | 22% | 7% |
| I regularly discuss the ethics of using generative AI with my students | Incorporators | 8% | 77% | 15% |
| | Non-Incorporators | 65% | 25% | 10% |
| I have caught students using generative AI against my stated syllabus policy | Incorporators | 58% | 34% | 8% |
| | Non-Incorporators | 90% | 10% | 0% |
| I have incorporated generative AI into my course assignments. | Incorporators | 46% | 48% | 6% |
| | Non-Incorporators | 76% | 24% | 0% |

TABLE XIV
GENAI USE: COMPARISON BETWEEN INCORPORATORS AND NON-INCORPORATORS IN SYLLABUS

| Perception Statement | Group | Response | | | |
|---|---|---|---|---|---|
| | | Strongly Disagree | Somewhat Disagree | Somewhat Agree | Strongly Agree |
| Generative AI will transform higher education classrooms for good. | Incorporators | 9% | 19% | 55% | 17% |
| | Non-Incorporators | 0% | 18% | 64% | 18% |
| Generative AI has or will influence how I design my curriculum. | Incorporators | 2% | 28% | 40% | 30% |
| | Non-Incorporators | 5% | 10% | 54% | 31% |
| Generative AI will transform the professional engineering workplace for good. | Incorporators | 2% | 13% | 57% | 28% |
| | Non-Incorporators | 3% | 15% | 53% | 23% |
| Generative AI has or will transform how I do course preparation and grading. | Incorporators | 0% | 13% | 53% | 34% |
| | Non-Incorporators | 3% | 26% | 45% | 26% |
| Generative AI has or will transform how I do my research. | Incorporators | 4% | 28% | 30% | 38% |
| | Non-Incorporators | 0% | 28% | 44% | 28% |

TABLE XV
PERCEPTIONS ABOUT GENAI: COMPARISON BETWEEN INCORPORATORS AND NON-INCORPORATORS IN SYLLABUS

These findings suggest that integrating and engaging with GenAI is critical to fostering positive interactions between instructors and students around this technology. However, there may very well be a gap in providing instructors with opportunities for exposure and instructional support, such as tutorials and case studies, which could further enhance their comfort and effectiveness in using GenAI in the classroom. Prior research suggests that schools are attempting to address these challenges [33]. Our analysis suggests that there may be reasons why instructors are resistant to this advice, as we discuss in the next section.

| GenAI Impact | Group | Response | | | |
|---|---|---|---|---|---|
| | | Negatively (Somewhat, Very) | Neither Negatively nor Positively | Positively (Somewhat, Very) | Does not apply |
| Critical thinking | Incorporators | 26% | 40% | 21% | 13% |
| | Non-Incorporators | 9% | 21% | 44% | 26% |
| Problem solving | Incorporators | 22% | 35% | 36% | 7% |
| | Non-Incorporators | 5% | 35% | 28% | 32% |
| Learning course concepts | Incorporators | 10% | 35% | 46% | 9% |
| | Non-Incorporators | 3% | 30% | 35% | 32% |
| Studying for a test | Incorporators | 9% | 41% | 20% | 30% |
| | Non-Incorporators | 5% | 25% | 22% | 48% |
| Writing prose | Incorporators | 21% | 28% | 42% | 9% |
| | Non-Incorporators | 17% | 13% | 28% | 42% |
| Proof Reading or Writing/generating Code | Incorporators | 13% | 15% | 55% | 17% |
| | Non-Incorporators | 8% | 8% | 40% | 44% |

TABLE XVI
GENAI IMPACT ON COURSE ACTIVITIES: COMPARISON BETWEEN INCORPORATORS AND NON-INCORPORATORS IN SYLLABUS

*B. GenAI Avoidance*

Instructors who choose not to engage with GenAI in the classroom may also be consciously avoiding it due to concerns about its legitimacy and potential erosion of learning. Our analysis highlights that a key challenge for instructors is the potential for GenAI to hinder genuine learning and compromise fair assessment practices. Issues such as the veracity of GenAI content have been widely discussed in the literature, as well as its impact on learning outcomes [25], [34], [35] and fair assessments [26], [27]. Luo [36] emphasizes that educators often perceive GenAI as a threat to the originality of students' work, associating its use with academic misconduct, such as plagiarism. However, our findings suggest that plagiarism itself is not among their primary concerns in the classroom. Rather, worries seem to be refocused on their ability to identify and assess counterfeit work.

That said, instructors are concerned about the role of plagiarism in academic research, particularly its impact on research integrity, more than they are about student plagiarism. It may be that their concerns intensify when plagiarism impacts the assessment of *their* work.

VI. LIMITATIONS AND RECOMMENDATIONS

This study captures educators' self-reported perceptions of GenAI use in syllabi and assignments to explore their alignment with perceptions of its value. While collecting quantitative data on actual syllabi and course materials might have provided additional insights, we have no reason to believe that educators misrepresented their use. Self-reported data is indeed appropriate for understanding personal perceptions and contextual experiences, which are central to this study's objectives.

A limitation of this study is that it draws on data from a small, preliminary survey conducted on educators before they attended a workshop on the use of GenAI for teaching and research. The workshop focused specifically on engineering education, and the majority of the respondents were faculty members in engineering or closely related to computing disciplines. This narrow scope limits the generalizability of our findings, as perspectives from faculty in other fields are not represented. Consequently, future research would benefit from a broader and more diverse sample to understand how GenAI adoption varies across academic fields and experience levels. Our research demonstrates a positive relationship between engagement with GenAI and both enthusiasm for and realistic perceptions of its use. We recommend that educators engage with GenAI, at a minimum, to develop greater awareness or cultivate a more nuanced understanding of its potential applications. While it is possible that enthusiasm for the tool may obscure its potential negative effects on skill development, as highlighted in the literature, we see no drawback in educators deepening their understanding of it. [32] provides guidance on best practices for incorporating GenAI into syllabi. Although our study was not explicitly designed to identify such best practices, the findings suggest that proactive engagement with GenAI is a promising approach.

VII. CONCLUSION

This study provides insights into how university-level engineering and computing educators perceive and integrate GenAI into their courses and research. By analyzing their responses through the lens of communication and integration practices, we identified how varying attitudes and levels of use shape teaching methods and perceptions. Our findings lay the groundwork for further exploration of GenAI's implications, particularly in the classroom, highlighting the need for ongoing support and resources to promote meaningful adoption and ethical engagement in both instruction and research.

VIII. ACKNOWLEDGMENTS

This work is partly supported by US NSF Awards 2319137, 1954556, and USDA/NIFA Award 2021-67021-35329. Any opinions, findings, and conclusions or recommendations expressed in this material are those of the authors and do not necessarily reflect the views of the funding agencies.